\documentclass[twocolumn,showpacs,aps,epsfig,nofootinbib]{revtex4}

\usepackage{graphicx}
\usepackage{epstopdf}
\usepackage{latexsym}
\usepackage{amssymb}

\usepackage[center]{subfigure}

\begin{document}

 \newcommand{\bq}{\begin{equation}}
 \newcommand{\eq}{\end{equation}}
 \newcommand{\bqn}{\begin{eqnarray}}
 \newcommand{\eqn}{\end{eqnarray}}
 \newcommand{\nb}{\nonumber}
 \newcommand{\lb}{\label}
\newcommand{\PRL}{Phys. Rev. Lett.}
\newcommand{\PL}{Phys. Lett.}
\newcommand{\PR}{Phys. Rev.}
\newcommand{\CQG}{Class. Quantum Grav.}

\title{Cosmology in nonrelativistic general covariant theory of gravity} 

\author{Anzhong Wang}
\email{anzhong_wang@baylor.edu}

\author{Yumei Wu}
\email{yumei_wu@baylor.edu}

\affiliation{GCAP-CASPER, Physics Department, Baylor
University, Waco, TX 76798-7316, USA}

\date{\today}

\begin{abstract}

Horava and  Melby-Thompson recently  proposed a new version of the Horava-Lifshitz theory of gravity,
in which the spin-0 graviton is eliminated by introducing a Newtonian pre-potential  $\varphi$ and a local 
$U(1)$ gauge field $A$. In this paper, we first derive the corresponding Hamiltonian, super-momentum 
constraints, the dynamical equations, and the equations for $\varphi$ and $A$, in the presence of matter 
fields. Then, we apply the theory to cosmology, and obtain the modified Friedmann  equation and the 
conservation law of energy, in addition to the equations for $\varphi$ and $A$. When the spatial curvature 
is different from zero, terms behaving  like dark radiation and stiff-fluid exist, from which, among other 
possibilities,  bouncing universe can be constructed. We also study linear perturbations of the FRW 
universe with any given spatial curvature $k$, and derive the most general formulas for scalar perturbations. 
The vector and tensor perturbations are the same as those recently given  by one of the present authors
[A. Wang, Phys. Rev. D{\bf 82}, 124063 (2010)] in the setup of Sotiriou, Visser and Weinfurtner. Applying 
these formulas to the Minkowski background, we have shown explicitly that the scalar and vector 
perturbations of the metric indeed vanish,  and the only remaining modes are the massless spin-2 gravitons.

\end{abstract}

\pacs{04.60.-m; 98.80.Cq; 98.80.-k; 98.80.Bp}

\maketitle

\section{Introduction}
\renewcommand{\theequation}{1.\arabic{equation}} \setcounter{equation}{0}

Recently, Horava proposed a quantum gravity theory \cite{Horava}, motivated by the Lifshitz 
theory in solid state physics \cite{Lifshitz}, for which the theory is often referred to as 
the Horava-Lifshitz (HL) theory. It  is non-relativistic and power-counting ultraviolet (UV)-renormalizable, 
and was expected to recover general relativity (GR) in the infrared (IR) limit. HL theory has attracted
a great deal of attention due to its several remarkable features, such as
  the divergence of its effective speed of light
in the UV, which could potentially resolve the horizon problem without 
invoking inflation  \cite{KK}. Scale-invariant super-horizon curvature 
perturbations can also be produced without inflation 
\cite{Muka,Piao,GWBR,SVW,WM,YKN,WWM,Wang}, and dark matter and dark energy 
can have their geometric origins \cite{Mukb,fR}. 
Furthermore, bouncing universe can be easily constructed due to the
high-order derivative terms of the spacetime curvature \cite{calcagni,brand,WWa}.
For detail, we refer readers to  \cite{HWW} and references therein.

The HL theory is based on the perspective 
that Lorentz symmetry should appear as an emergent symmetry at long distances, but can be fundamentally 
absent at high energies \cite{Pav}. With this in mind, 
Horava considered systems whose scaling at short 
distances exhibits a strong anisotropy between space and time, 
\bq
\lb{1.1}
{\bf x} \rightarrow \ell {\bf x}, \;\;\;  t \rightarrow \ell^{z} t.
\eq
In $(d+1)$-dimensional spacetimes, in order for the theory to be power-counting 
renormalizable,    it requires $z \ge d$. At low energies, the theory is expected to  flow 
to $z = 1$, whereby the Lorentz invariance is ``accidentally restored." 
Such an anisotropy between  time and space can be easily realized when one writes the metric
in the Arnowitt-Deser-Misner  (ADM) form  \cite{ADM},
 \bqn
 \lb{1.2}
ds^{2} &=& - N^{2}c^{2}dt^{2} + g_{ij}\left(dx^{i} + N^{i}dt\right)
     \left(dx^{j} + N^{j}dt\right), \nb\\
     & & ~~~~~~~~~~~~~~~~~~~~~~~~~~~~~~  (i, \; j = 1, 2, 3).~~~
 \eqn
 Under the rescaling (\ref{1.1})  
 with   $z = d = 3$, a condition we shall assume 
 in the rest of this paper,  the dynamical variables $N, \; N^{i}$ and $g_{ij}$ scale,
 respectively,  as, 
 \bq
 \lb{1.3}
  N \rightarrow  N ,\;  N^{i}
\rightarrow {\ell}^{-2} N^{i},\; g_{ij} \rightarrow g_{ij}.
 \eq
 
The gauge symmetry of the system are the  foliation-preserving 
diffeomorphisms Diff($M, \; {\cal{F}}$),  
\bq
\lb{1.4}
\tilde{t} = t - f(t),\; \;\; \tilde{x}^{i}  =  {x}^{i}  - \zeta^{i}(t, {\bf x}),
\eq
for which the dynamical variables change as
\bqn
\lb{1.5}
\delta{g}_{ij} &=& \nabla_{i}\zeta_{j} + \nabla_{j}\zeta_{i} + f\dot{g}_{ij},\nb\\
\delta{N}_{i} &=& N_{k}\nabla_{i}\zeta^{k} + \zeta^{k}\nabla_{k}N_{i}  + g_{ik}\dot{\zeta}^{k}
+ \dot{N}_{i}f + N_{i}\dot{f}, \nb\\
\delta{N} &=& \zeta^{k}\nabla_{k}N + \dot{N}f + N\dot{f},
\eqn
where $\dot{f} \equiv df/dt$,  $\nabla_{i}$ denotes the covariant 
derivative with respect to the 3-metric $g_{ij}$, and  $N_{i} = g_{ik}N^{k}$, etc. From these expressions one can see that the
lapse function $N$ and the shift vector $N_{i}$ play the role of gauge fields of the Diff($M, \; {\cal{F}}$)
symmetry. Therefore, it is natural to assume that $N$ and $N_{i}$ inherit the same dependence on 
spacetime as the corresponding generators, in addition to the fact that the dynamical variables $g_{ij}$
should  in general depend on both time and space, that is, 
\bq
\lb{1.6}
N = N(t), \;\;\; N_{i} = N_{i}(t, x),\;\;\; g_{ij} = g_{ij}(t, x),
\eq
which is   clearly preserved by the Diff($M, \; {\cal{F}}$), and often referred to as the projectability condition.

Due to the restricted diffeomorphisms (\ref{1.4}), one more degree of freedom appears
in the gravitational sector - a spin-0 graviton. This is potentially dangerous, and needs to decouple  
in the IR regime, in order to be consistent with observations.   Unfortunately,  it was shown that this
might not be the case. In particular,  the spin-0 mode is not stable 
 in the original version of the HL theory  \cite{Horava} as well as in the Sotiriou, Visser 
and Weinfurtner (SVW) generalization \cite{SVW,WM}. Note that in both of these two versions, it was 
all assumed the projectability condition. In addition, these
instabilities are all found in the Minkowski background. Recently, it was found that the de Sitter
spacetime is stable in the SVW setup \cite{HWW}. So, one can take the latter 
as its legitimate  background, similar to what happened in the massive gravity \cite{MGs}. However,
the strong coupling problem still exists \cite{KA,WWb}, although it might be   circumvented 
by the Vainshtein mechanism \cite{Vain}, as recently showed in the spherical   \cite{Mukc} and
  cosmological \cite{WWb} cases.

On the other hand, giving up the projectability  condition, that is, assuming that the lapse function $N$
depends on both time  and spatial coordinates,
Blas, Pujolas and Sibiryakov (BPS) \cite{BPSb} found that inclusion of  terms made of $a_{i}$,
 \bq
 \lb{1.7}
 a_{i} = \partial_{i}\ln(N),
 \eq
can cure the instability of the Minkowski spacetime.  By properly choosing the coupling constants, 
the strong coupling problem \cite{CNPS,BPSa,PS,KP} can be also addressed \cite{BPSc}.  
However, a  price to pay     is the enormous number of independent 
coupling constants:  only the sixth-order derivative terms 
in the potential are more than 60  \cite{KP}. It should be also noted that
giving  up the projectability condition often causes the theory to suffer the inconsistence 
problem \cite{LP}.   Kluson  recently showed that the Hamiltonian formalism of the BPS model 
is very rich, and the corresponding  algebra of constraints is well-defined   \cite{Kluson}. 

To cure the instability problem, another very attractive way is to eliminate the spin-0 graviton from the theory, 
so that the resulting one has as many generators per spacetime
point as  GR does.  This is done recently  by Horava and Melby-Thompson (HMT) \cite{HMT}  (with the assumption
of the projectability condition (\ref{1.6})) by   extending 
the foliation-preserving-diffeomorphisms, Diff($M, \; {\cal{F}}$), to include  a local $U(1)$ symmetry, 
\bq
\lb{symmetry}
 U(1) \ltimes {\mbox{Diff}}(M, \; {\cal{F}}).
 \eq
Effectively, the spatial diffeomorphism symmetries of GR are kept intact, but its time reparametrization symmetry is linearized
and the corresponding  algebra is contracted to a local gauge symmetry \cite{TH}. The restoration of general covariance, 
characterized by  Eq.(\ref{symmetry}),  nicely maintains the special status of time,  so that the anisotropic scaling (\ref{1.1})
with $z > 1$ can still be  realized. 

A remarkable by-production of this ``non-relativistic general covariant" setup is that it forces the coupling constant
$\lambda$, introduced originally  to characterize the deviation of the kinetic part of the action from GR \cite{Horava},
to take exactly  its relativistic value $\lambda = 1$. Note that in GR the spacetime diffeomorphism symmetry, 
Diff($M$),  
\bq
\lb{1.8}
\tilde{x}^{\mu} = {x}^{\mu}  - \zeta^{\mu}(t, {\bf x}),\; (\mu = 0, 1, 2, 3)
\eq
also forces $\lambda = 1$ and protects this value from quantum corrections. 

At short distances, the theory exhibits a high anisotropy between time and space.
As a result, the UV behavior of the theory is dramatically imporved. At long distances, the theory is driven to an IR regime,
where it shares many features with GR. In particular, under the influences of the relevant terms, the scaling is naturally isotropic
with the relativistic value $z = 1$. Moreover, since the extended symmetry forces $\lambda = 1$, in the IR limit the action will be
dominated exactly by the Einstein-Hilbert terms in the ADM decomposition \cite{ADM}.

In this paper,  we  investigate this new version of the HL theory. Specifically, in Sec. II we  first give a brief review of
it, and then derive the   corresponding Hamiltonian, super-momentum constraints, the dynamical equations, and the field equations for 
the Newtonian pre-potential   $\varphi$ and the $U(1)$  gauge field $A$, in the presence of matter fields. The potential used in this
paper is the one constructed by SVW \cite{SVW}, which represents the most general form, subjected to the assumptions that it respects the parity and
its highest order of the spatial derivatives is six, the minimal requirement to have the theory be power-counting renormalizable \cite{Horava}.
In Sec. III,  we apply the theory to cosmology, and obtain the modified Friedmann equation and conservation law of energy,  in addition to
 the  equations for  $\varphi$ and $A$. In Sec. IV we  study the linear perturbations of the FRW universe with any given spatial curvature 
 $k$, and present the general formulas for scalar perturbations. The vector and tensor perturbations  are the same as those given by
 one of the current authors in \cite{Wang} in the SVW setup \cite{SVW}, because the gauge field and the Newtonian pre-potential have 
 no contributions to these parts. Applying these formulas to the Minkowski background in Sec. V, we show explicitly that the scalar and
 vector perturbations of the metric vanish identically. The only non-vanishing dynamical variables are the traceless and divergence-free tensor
 $H_{ij}$, which describes the massless spin-2 graviton, a situation that is precisely the same as in GR.  These results are consistent with 
the ones obtained earlier  in \cite{HMT}.  In Sec. VI, we present our main conclusions.

\section{Non-relativisitc general covariant HL theory}

\renewcommand{\theequation}{2.\arabic{equation}} \setcounter{equation}{0}


In   order to limit the spin-0 graviton, HMT   introduced two new fields, the $U(1)$ gauge field $A$ 
and the Newtonian pre-potential $\varphi$, where in general both of them depend on  space and time, 
\bq
\lb{2.1}
A = A\left(t, x^{k}\right),\;\;\;
\varphi= \varphi\left(t, x^{k}\right).
\eq
Note that  the notations used  in this paper are   slightly different from those adopted in \cite{HMT}  \footnote{In particular, we
have $\varphi = - \nu^{HMT},\; K_{ij} = - K_{ij}^{HMT},\; \Lambda_{g} = \Omega^{HMT},\;
{\cal{G}}_{ij} = \Theta_{ij}^{HMT}$, where quantities with super indice ``HMT" are the ones used in \cite{HMT}.}. Under the Diff($M, \; {\cal{F}}$),
these fields transfer as,
\bqn
\lb{2.2}
\delta{A} &=& \zeta^{i}\partial_{i}A + \dot{f}A  + f\dot{A},\nb\\
\delta \varphi &=&  f \dot{\varphi} + \zeta^{i}\partial_{i}\varphi,
\eqn
while under the local $U(1)$,  they, together with $g_{ij}$,
 transfer as
\bqn
\lb{2.3}
\delta_{\alpha}A &=&\dot{\alpha} - N^{i}\nabla_{i}\alpha,\;\;\;
\delta_{\alpha}\varphi = - \alpha,\nb\\ 
\delta_{\alpha}N_{i} &=& N\nabla_{i}\alpha,\;\;\;
\delta_{\alpha}g_{ij} = 0 = \delta_{\alpha}{N},
\eqn
where $\alpha$ is   the generator  of the local $U(1)$ gauge symmetry. 
For the detail, we refer readers to \cite{HMT}. 

The total action is given by,
 \bqn \lb{2.4}
S &=& \zeta^2\int dt d^{3}x N \sqrt{g} \Big({\cal{L}}_{K} -
{\cal{L}}_{{V}} +  {\cal{L}}_{{\varphi}} +  {\cal{L}}_{{A}} \nb\\
& & ~~~~~~~~~~~~~~~~~~~~~~ \left. +\frac{1}{\zeta^{2}} {\cal{L}}_{M} \right),
 \eqn
where $g={\rm det}\,g_{ij}$, and
 \bqn \lb{2.5}
{\cal{L}}_{K} &=& K_{ij}K^{ij} -   K^{2},\nb\\
{\cal{L}}_{\varphi} &=&\varphi {\cal{G}}^{ij} \Big(2K_{ij} + \nabla_{i}\nabla_{j}\varphi\Big),\nb\\
{\cal{L}}_{A} &=&\frac{A}{N}\Big(2\Lambda_{g} - R\Big).
 \eqn
Here   the coupling constant $\Lambda_{g}$, acting like a 3-dimensional cosmological
constant, has the dimension of (length)$^{-2}$. The 
Ricci and Riemann terms all refer to the three-metric $g_{ij}$,
 $K_{ij}$ is the extrinsic curvature, and ${\cal{G}}_{ij}$ is the 3-dimensional ``generalized"
Einstein tensor, defined, respectively, by
 \bqn \lb{2.6}
K_{ij} &=& \frac{1}{2N}\left(- \dot{g}_{ij} + \nabla_{i}N_{j} +
\nabla_{j}N_{i}\right),\nb\\
{\cal{G}}_{ij} &=& R_{ij} - \frac{1}{2}g_{ij}R + \Lambda_{g} g_{ij}.
 \eqn
${\cal{L}}_{M}$ is the
matter Lagrangian density, which in general is a function of all the dynamical variables,
$U(1)$ gauge field, and the Newtonian prepotential, 
\bq
\lb{2.6a}
{\cal{L}}_{M} = {\cal{L}}_{M}\big(N, \; N_{i}, \; g_{ij}, \; \varphi,\; A; \; \chi\big),
\eq
where $\chi$ denotes collectively the matter fields. ${\cal{L}}_{{V}}$ is an arbitrary Diff($\Sigma$)-invariant local scalar functional
built out of the spatial metric, its Riemann tensor and spatial covariant derivatives, without the use of time derivatives. In the original 
approach of Horava \cite{Horava}, the detailed balance condition was imposed, in order to limit the number of the coupling constants. 
With this condition, 
${\cal{L}}_{{V}}$ takes the simple form,
\bq
\lb{DBC}
{\cal{L}}_{{V}} = w^{2}C_{ij}C^{ij},
\eq
where $w$ is a coupling constant, and $C_{ij}$ denotes the Cotton tensor, defined by
\bq
\lb{DBCa}
C^{ij} = \epsilon^{ikl}\nabla_{k}\Big(R^{j}_{l} - \frac{1}{4}R\delta^{j}_{l}\Big).
\eq
In \cite{SVW}, by assuming that the highest order derivatives are six and that  the theory  respects 
the parity, SVW constructed the most general form of  ${\cal{L}}_{{V}}$, given by
 \bqn \lb{2.5a} 
{\cal{L}}_{{V}} &=& \zeta^{2}g_{0}  + g_{1} R + \frac{1}{\zeta^{2}}
\left(g_{2}R^{2} +  g_{3}  R_{ij}R^{ij}\right)\nb\\
& & + \frac{1}{\zeta^{4}} \left(g_{4}R^{3} +  g_{5}  R\;
R_{ij}R^{ij}
+   g_{6}  R^{i}_{j} R^{j}_{k} R^{k}_{i} \right)\nb\\
& & + \frac{1}{\zeta^{4}} \left[g_{7}R\nabla^{2}R +  g_{8}
\left(\nabla_{i}R_{jk}\right)
\left(\nabla^{i}R^{jk}\right)\right],  ~~~~
 \eqn 
 where the coupling  constants $ g_{s}\, (s=0, 1, 2,\dots 8)$  are all dimensionless. The relativistic limit in the IR
 requires $g_{1} = -1$ and $\zeta^2 = 1/(16\pi G)$ \cite{SVW}. 
In this paper, we shall be concerned only with this potential, and
  our formulas to be obtained below can be easily generalized to other forms of the potential, including the
 one given by Eq.(\ref{DBC}), and the $f(R)$ term studied in \cite{fR}.

Variation with respect to the lapse function $N(t)$  yields the
Hamiltonian constraint,
 \bq \lb{eq1}
\int{ d^{3}x\sqrt{g}\left({\cal{L}}_{K} + {\cal{L}}_{{V}} - \varphi {\cal{G}}^{ij}\nabla_{i}\nabla_{j}\varphi\right)}
= 8\pi G \int d^{3}x {\sqrt{g}\, J^{t}},
 \eq
where
 \bq \lb{eq1a}
J^{t} = 2 \frac{\delta\left(N{\cal{L}}_{M}\right)}{\delta N}.
 \eq

Variation with respect to the shift $N^{i}$ yields the
super-momentum constraint,
 \bq \lb{eq2}
\nabla_{j}\Big(\pi^{ij} - \varphi  {\cal{G}}^{ij}\Big) = 8\pi G J^{i},
 \eq
where the super-momentum $\pi^{ij} $ and matter current $J^{i}$
are defined as
 \bqn \lb{eq2a}
\pi^{ij} &\equiv& \frac{\delta{(N\cal{L}}_{K})}{\delta\dot{g}_{ij}}
 = - K^{ij} +  K g^{ij},\nb\\
J^{i} &\equiv& - N\frac{\delta{\cal{L}}_{M}}{\delta N_{i}}.
 \eqn
Similarly, variations of the action with respect to $\varphi$ and $A$ yield, 
\bqn
\lb{eq4a}
& & {\cal{G}}^{ij} \Big(K_{ij} + \nabla_{i}\nabla_{j}\varphi\Big) = 8\pi G J_{\varphi},\\
\lb{eq4b}
& & R = 2\Lambda_{g} +   8\pi G J_{A},
\eqn
where
\bq
\lb{eq5}
J_{\varphi} \equiv - \frac{\delta{\cal{L}}_{M}}{\delta\varphi},\;\;\;
J_{A} \equiv 2 \frac{\delta\left(N{\cal{L}}_{M}\right)}{\delta{A}}.
\eq
On the other hand, variation with respect to $g_{ij}$ leads to the
dynamical equations,
 \bqn \lb{eq3}
&&
\frac{1}{N\sqrt{g}}\left[\sqrt{g}\left(\pi^{ij} - \varphi {\cal{G}}^{ij}\right)\right]_{,t} 
= -2\left(K^{2}\right)^{ij}+2K K^{ij}
\nb\\
& &  ~~~~~ + \frac{1}{N}\nabla_{k}\left[N^k \pi^{ij}-2\pi^{k(i}N^{j)}\right]\nb\\
& & ~~~~~
+  \frac{1}{2} \left({\cal{L}}_{K} + {\cal{L}}_{\varphi} + {\cal{L}}_{A}\right) g^{ij} \nb\\
& &  ~~~~~    + F^{ij} + F_{\varphi}^{ij} +  F_{A}^{ij} + 8\pi G \tau^{ij},
 \eqn
where $\left(K^{2}\right)^{ij} \equiv K^{il}K_{l}^{j},\; f_{(ij)}
\equiv \left(f_{ij} + f_{ji}\right)/2$, and
 \bqn
\lb{eq3a} 
F^{ij} &\equiv&
\frac{1}{\sqrt{g}}\frac{\delta\left(-\sqrt{g}
{\cal{L}}_{V}\right)}{\delta{g}_{ij}}
 = \sum^{8}_{s=0}{g_{s} \zeta^{n_{s}}
 \left(F_{s}\right)^{ij} },\nb\\
F_{\varphi}^{ij} &=&  \sum^{3}_{n=1}{F_{(\varphi, n)}^{ij}},\nb\\
F_{A}^{ij} &=& \frac{1}{N}\left[AR^{ij} - \Big(\nabla^{i}\nabla^{j} - g^{ij}\nabla^{2}\Big)A\right],\nb\\ 
 \eqn
with 
$n_{s} =(2, 0, -2, -2, -4, -4, -4, -4,-4)$. The
stress 3-tensor $\tau^{ij}$ is defined as
 \bq \label{tau}
\tau^{ij} = {2\over \sqrt{g}}{\delta \left(\sqrt{g}
 {\cal{L}}_{M}\right)\over \delta{g}_{ij}},
 \eq
and the geometric 3-tensors $ \left(F_{s}\right)_{ij}$ and $F_{(\varphi, n)}^{ij}$ are defined
as,
  \bqn \lb{eq3b}
\left(F_{0}\right)_{ij} &=& - \frac{1}{2}g_{ij},\nb\\
\left(F_{1}\right)_{ij} &=& R_{ij}- \frac{1}{2}Rg_{ij},\nb\\
\left(F_{2}\right)_{ij} &=& 2\left(R_{ij} -
\nabla_{i}\nabla_{j}\right)R
-  \frac{1}{2}g_{ij} \left(R - 4\nabla^{2}\right)R,\nb\\
\left(F_{3}\right)_{ij} &=& \nabla^{2}R_{ij} - \left(\nabla_{i}
\nabla_{j} - 3R_{ij}\right)R - 4\left(R^{2}\right)_{ij}\nb\\
& & +  \frac{1}{2}g_{ij}\left( 3 R_{kl}R^{kl} + \nabla^{2}R
- 2R^{2}\right),\nb\\
\left(F_{4}\right)_{ij} &=& 3 \left(R_{ij} -
\nabla_{i}\nabla_{j}\right)R^{2}
 -  \frac{1}{2}g_{ij}\left(R  - 6 \nabla^{2}\right)R^{2},\nb\\
 \left(F_{5}\right)_{ij} &=&  \left(R_{ij} + \nabla_{i}\nabla_{j}
 \right) \left(R_{kl}R^{kl}\right)
 + 2R\left(R^{2}\right)_{ij} \nb\\
& & + \nabla^{2}\left(RR_{ij}\right) - \nabla^{k}\left[\nabla_{i}
\left(RR_{jk}\right) +\nabla_{j}\left(RR_{ik}\right)\right]\nb\\
& &  -  \frac{1}{2}g_{ij}\left[\left(R - 2 \nabla^{2}\right)
\left(R_{kl}R^{kl}\right)\right.\nb\\
& & \left.
- 2\nabla_{k}\nabla_{l}\left(RR^{kl}\right)\right],\nb\\
\left(F_{6}\right)_{ij} &=&  3\left(R^{3}\right)_{ij}  +
\frac{3}{2}
\left[\nabla^{2}\left(R^{2}\right)_{ij} \right.\nb\\
 & & \left.
 - \nabla^{k}\left(\nabla_{i}\left(R^{2}\right)_{jk} + \nabla_{j}
 \left(R^{2}\right)_{ik}\right)\right]\nb\\
 & &    -  \frac{1}{2}g_{ij}\left[R^{k}_{l}R^{l}_{m}R^{m}_{k} -
 3\nabla_{k}\nabla_{l}\left(R^{2}\right)^{kl}\right],\nb\\
 \left(F_{7}\right)_{ij} &=&  2 \nabla_{i}\nabla_{j}
 \left(\nabla^{2}R\right) - 2\left(\nabla^{2}R\right)R_{ij}\nb\\
 & &    + \left(\nabla_{i}R\right)\left(\nabla_{j}R\right)
  -  \frac{1}{2}g_{ij}\left[\left(\nabla{R}\right)^{2} +
  4 \nabla^{4}R\right],\nb\\
\left(F_{8}\right)_{ij} &=&  \nabla^{4}R_{ij} -
\nabla_{k}\left(\nabla_{i}\nabla^{2} R^{k}_{j}
                            + \nabla_{j}\nabla^{2} R^{k}_{i}
                            \right)\nb\\
& & - \left(\nabla_{i}R^{k}_{l}\right)
\left(\nabla_{j}R^{l}_{k}\right)
       - 2 \left(\nabla^{k}R^{l}_{i}\right) \left(\nabla_{k}R_{jl}
       \right)\nb\\
& &    -  \frac{1}{2}g_{ij}\left[\left(\nabla_{k}R_{lm}\right)^{2}
        -
        2\left(\nabla_{k}\nabla_{l}\nabla^{2}R^{kl}\right)\right],\\
  \lb{eq3c}
F_{(\varphi, 1)}^{ij} &=& \frac{1}{2}\varphi\left\{\Big(2K + \nabla^{2}\varphi\Big) R^{ij}  
- 2 \Big(2K^{j}_{k} + \nabla^{j} \nabla_{k}\varphi\Big) R^{ik} \right.\nb\\
& & ~~~~~ - 2 \Big(2K^{i}_{k} + \nabla^{i} \nabla_{k}\varphi\Big) R^{jk}\nb\\
& &~~~~~\left. 
- \Big(2\Lambda_{g} - R\Big) \Big(2K^{ij} + \nabla^{i} \nabla^{j}\varphi\Big)\right\},\nb\\
F_{(\varphi, 2)}^{ij} &=& \frac{1}{2}\nabla_{k}\left\{\varphi{\cal{G}}^{ik}  
\Big(\frac{2N^{j}}{N} + \nabla^{j}\varphi\Big) \right. \nb\\
& & \left.
+ \varphi{\cal{G}}^{jk}  \Big(\frac{2N^{i}}{N} + \nabla^{i}\varphi\Big) 
-  \varphi{\cal{G}}^{ij}  \Big(\frac{2N^{k}}{N} + \nabla^{k}\varphi\Big)\right\}, \nb\\   
F_{(\varphi, 3)}^{ij} &=& \frac{1}{2}\left\{2\nabla_{k} \nabla^{(i}f^{j) k}_{\varphi}  
- \nabla^{2}f_{\varphi}^{ij}   - \left(\nabla_{k}\nabla_{l}f^{kl}_{\varphi}\right)g^{ij}\right\},\nb\\
\eqn
where
\bqn
\lb{eq3d}
f_{\varphi}^{ij} &=& \varphi\left\{\Big(2K^{ij} + \nabla^{i}\nabla^{j}\varphi\Big) 
- \frac{1}{2} \Big(2K + \nabla^{2}\varphi\Big)g^{ij}\right\}.\nb\\
\eqn

The matter quantities $(J^{t}, \; J^{i},\; J_{\varphi},\; J_{A},\; \tau^{ij})$ satisfy the
conservation laws,
 \bqn \lb{eq5a} & &
 \int d^{3}x \sqrt{g} { \left[ \dot{g}_{kl}\tau^{kl} -
 \frac{1}{\sqrt{g}}\left(\sqrt{g}J^{t}\right)_{, t}  
 +   \frac{2N_{k}}  {N\sqrt{g}}\left(\sqrt{g}J^{k}\right)_{,t}
  \right.  }   \nb\\
 & &  ~~~~~~~~~~~~~~ \left.   - 2\dot{\varphi}J_{\varphi} -  \frac{A} {N\sqrt{g}}\left(\sqrt{g}J_{A}\right)_{,t}
 \right] = 0,\\
\lb{eq5b} & & \nabla^{k}\tau_{ik} -
\frac{1}{N\sqrt{g}}\left(\sqrt{g}J_{i}\right)_{,t}  - \frac{J^{k}}{N}\left(\nabla_{k}N_{i}
- \nabla_{i}N_{k}\right)   \nb\\
& & \;\;\;\;\;\;\;\;\;\;\;- \frac{N_{i}}{N}\nabla_{k}J^{k} + J_{\varphi} \nabla_{i}\varphi - \frac{J_{A}}{2N} \nabla_{i}A
 = 0.
\eqn

\section{Cosmological Models}

\renewcommand{\theequation}{3.\arabic{equation}} \setcounter{equation}{0}

The homogeneous and isotropic universe is described by,
\bq
\lb{3.1}
N = 1,\;\; N_{i} = 0,\;\; 
g_{ij} = a^{2}(t)\gamma_{ij},
\eq
where 
\bq
\lb{3.2}
\gamma_{ij}=\frac{\delta_{ij}}{\left(1 + \frac{1}{4}kr^{2}\right)^{2}},
\eq
with $r^{2} \equiv x^2 + y^2 + z^2$ and $k = 0, \pm 1$. Using the $U(1)$ gauge, on the other hand, we can set
\bq
\lb{3.2a}
\varphi = 0,
\eq
without loss of generality. 
Then, we find that
\bq
\lb{3.3}
K_{ij} = - a^{2}H \gamma_{ij}, \;\;\; R_{ij} = 2k\gamma_{ij}, 
\eq
where $H = \dot{a}/a$. Thus,  we obtain
 \bqn
  \lb{3.4}
{\cal{L}}_{K} &=&  -6 H^{2},\;\; {\cal{L}}_{\varphi} = 0, \nb\\
 {\cal{L}}_{A} &=& 2A \left(\Lambda_{g} - \frac{3k}{a^{2}}\right),\nb\\
{\cal{L}}_{V} &=& 2\Lambda + \frac{6kg_1}{a^{2}} +
\frac{12\beta_1k^{2}}{a^{4}}  + \frac{24\beta_2k^{3}}{a^{6}},
 \eqn
where $\Lambda \equiv \zeta^{2} g_{0}/2$, and 
\bq
\lb{3.5}
\beta_1 \equiv \frac{3g_{2} + g_{3}} {\zeta^{2}},\;\;\; 
\beta_2 \equiv \frac{9g_{4} + 3g_{5} + g_{6}}{\zeta^{4}}.
\eq
 The matter components are
 \bq\label{mq}
{J}^t=-2\rho,~~ {J}^i=0,~~ \tau_{ij} =  p\,
g_{ij},
 \eq
where $\rho$ and $ p$ are the total density and pressure of the matter fields.
Then the Hamiltonian constraint (\ref{eq1}) reduces to the
super-Hamiltonian constraint, 
$$
{\cal{L}}_{K}(t) +
{\cal{L}}_{V}(t) =  8\pi GJ^{t}(t),
$$
which leads to the
modified Friedmann equation,
 \bq \lb{3.6a}
H^{2} - \frac{g_{1}k}{a^{2}} =
\frac{8\pi G}{3} \rho+ \frac{\Lambda}{3} 
+ \frac{2\beta_1k^{2}}{a^{4}} + \frac{4\beta_2k^{3}}{a^{6}}.
 \eq
From Eqs.~(\ref{eq2a}) and (\ref{eq3a}) we also find that
 \bqn \lb{3.7}
F^{ij} &=& \left(-\Lambda - \frac{g_{1}k}{a^{2}} +
\frac{2\beta_{1}k^{2}}{a^{4}}  + \frac{12\beta_{2}k^{3}}{a^{6}}
\right){g}^{ij},\nb\\
 \pi ^{ij}  &=& -2H g^{ij}.
 \eqn
Hence,  the dynamical equation (\ref{eq3}) reduces to \cite{WM}
 \bqn
 \lb{3.6b}
\frac{\ddot{a}}{a} &=&  - {4\pi G\over
3}(\rho+3 p)+ {1\over3} \Lambda - \frac{A}{2}\left(\Lambda_{g} - \frac{k}{a^{2}}\right)\nb\\
  & &
- \frac{2\beta_{1}k^{2}}{a^{4}}  - \frac{8\beta_{2}k^{3}}{a^{6}}.
 \eqn
Similar to GR, the super-momentum constraint
(\ref{eq2}) is then satisfied identically, since $ J^{i} = 0$
and, from Eq.~(\ref{3.5}),  $\vec\nabla_j \pi^{ij}\equiv
\pi^{ij}{}{}_{|j} = 0$, where $\vec\nabla_i$ denotes the
covariant derivative with respect to $\gamma_{ij}$.
Using Eqs.~(\ref{3.6a}) and (\ref{3.6b}), it follows that in the FRW
background the matter satisfies the conservation law,
 \bq \lb{3.6e}
\dot{{\rho}} + 3H \left(\rho + p \right) = A J_{\varphi}.
 \eq
 Thus,  due to the interaction between the gauge field and the fluid, its energy in general  is not conserved. 
 
 On the other hand, Eqs.(\ref{eq4a}) and (\ref{eq4b}) yield, respectively,
 \bqn
 \lb{3.8a}
 & & H\left(\Lambda_{g} - \frac{k}{a^{2}}\right) = - \frac{8\pi G}{3} J_{\varphi},\\
 \lb{3.8b}
 & & \frac{3k}{a^{2}} -  \Lambda_{g}=  4\pi G  J_{A}.
 \eqn
 
 When matter is not present, we have $J_{\varphi} = 0 = J_{A}$. Then, Eqs.(\ref{3.8a}) and (\ref{3.8b})
 implies that $\Lambda_{g} = 0 = k$, while Eq.(\ref{3.6a}) yields $a(t) = e^{H t}$, where $H \equiv
 \sqrt{\Lambda/3}$, which is the de Sitter spacetime. It is interesting to note that the de Sitter space
 can be also obtained from $\rho = p = k = 0$ and $J_{A} = -\Lambda_{g}/(4\pi G),\;
 J_{\varphi} = -3H\Lambda_{g}/(8\pi G)$. 
 

It should be noted that the energy conservation (\ref{3.6e}) can be also obtained from
Eq.~(\ref{eq5a}), while the conservation law of momentum, Eq.~(\ref{eq5b}),  is satisfied identically.
When $\beta_1 \beta_2 \not=0$, the corresponding 
 terms act like a dark radiation and a stiff-fluid, respectively. Due to the presence of these terms, one can
 easily construct bouncing universe in the early epoch of the universe \cite{calcagni,brand,WWa}.

In addition, in deriving Eq.~(\ref{3.6a}) we followed the usual assumption that
the whole FRW universe is homogeneous and isotropic. In
\cite{Mukb}, it was argued that such an assumption might be too
strong. If one relaxes the assumption and requires that only the
observed patch of our universe is homogeneous and isotropic, one
can introduce the notion of ``dark matter as an integration
constant" of the Hamiltonian constraint~(\ref{eq1}): $
\rho(t)$ in Eqs.~(\ref{3.6a}) and (\ref{3.6e}) can be replaced by
$\rho(t)+ {\cal E}(t)$ in the observable patch, where ${\cal
E}(t)=\mbox{const}/a^3$ in the IR limit
\cite{Mukb,Kob}. Beyond the observable patch, ${\cal
E}$ is necessarily inhomogeneous. In order to analyze
perturbations on an FRW background, one needs to restrict the
perturbations to the observable patch, which then raises issues
about matching across the boundary of the observable patch. In our
approach, the background is a homogeneous FRW spacetime, so that
${\cal E}=0$ in the background.

\section{Cosmological Scalar Perturbations }

\renewcommand{\theequation}{4.\arabic{equation}} \setcounter{equation}{0}

In this section, we consider linear scalar perturbation of the FRW universe studied in the last section.
We shall closely follow \cite{WM} and use the notations adopted there without further explanations. However, in order
to have the present paper as independent as possible, it is difficult to avoid repeating the same materials,
although we shall try to limit it to its minimum. 

In the quasi-longitudinal gauge \cite{WM}, 
\bq
\lb{gauge}
\phi = 0 = E,
\eq
the metric scalar perturbations
are given by
 \bqn \lb{4.1}
ds^{2} &=& a^{2}\left[- d\eta^{2} + 2 {B}_{|i}dx^{i}d\eta +
\left(1 - 2 \psi\right)\gamma_{ij}dx^{i}dx^{j}\right].\nb\\
 \eqn
 The gauge-invaraint quantities are now given by
 \bq
 \lb{4.1a}
 \Phi = {\cal{H}}B + B',\;\;
 \Psi = \psi - {\cal{H}}B,
 \eq
 where $ {\cal{H}} \equiv a'/a$ and a prime denotes the ordinary derivative with respect to $\eta$.
Using the $U(1)$ gauge, we can further set 
\bq
\lb{gaugea}
\delta\varphi = 0,
\eq
so that ${\cal{L}}_{\varphi} = 0$. 
 Following \cite{WM}, we use quantities with over-bars as the ones calculated in the
 background.  Then   we find that
 \bqn \lb{4.2}
K_{ij} &=& -a{\cal{H}}\gamma_{ij}  +   a\left[B_{|ij}
 +  \left(\psi' + 2{\cal{H}}\psi\right)\gamma_{ij}\right],\nb\\
 R_{ij} &=& 2k\gamma_{ij}   +  \psi_{|ij} + \vec{\nabla}^{2}\psi \gamma_{ij},\nb\\
 {\cal{L}}_{K} &=& - \frac{6{\cal{H}}^{2}}{a^{2}} +
  \frac{4{\cal{H}}}{a^{2}}   \left(\vec\nabla ^{2}B  + 3 \psi'\right),\nb\\
{\cal{L}}_{A} &=&\bar{ {\cal{L}}}_{A}  + \frac{2 }{a}\Bigg[\Big(\Lambda_{g} - \frac{3k}{a^{2}}\Big)\delta{A}
  - \frac{2\bar{A}}{a^{2}}\Big(\vec{\nabla}^{2}\psi + 3k\psi\Big)\Bigg],\nb\\
 {\cal{L}}_{V} &=&  \bar{\cal{L}}_{V}
  + \frac{4 }{a^{2}}\left(g_{1} + \frac{4\beta_1k}{a^{2}}\right)
  \left(\vec\nabla ^{2} + 3k\right)\psi \nb\\
 & &~ +  \frac{48 \beta_2k^{2}}{a^{6}} \left(\vec\nabla ^{2}
+ 3k\right)\psi  \nb\\ &&~  +  \frac{24  g_{7}
k}{\zeta^{4}a^{6}}\vec\nabla ^{2}
 \left(\vec\nabla ^{2} + 3k\right)\psi,
 \eqn
 where $ \bar{\cal{L}}_{V}$ denotes the potential of the background given by Eq.(\ref{3.4}),
 and $B_{|i} \equiv \vec{\nabla}_{i}$, with  $\vec{\nabla}_{i}$ being the covariant derivative with 
 respect to $\gamma_{ij}$ and $\vec\nabla ^{2} \equiv \gamma^{ij} \vec\nabla_{i}\vec\nabla_{j}$.
 
To first-order the Hamiltonian constraint (\ref{eq1}) takes the form,
 \bqn \lb{4.4}
& & \int \sqrt{\gamma}d^{3}x\Bigg[\left(\vec\nabla^2+3k\right)\psi
- \frac{(2-3\xi){\cal H}}{2}
\left(\vec\nabla^2 B + 3\psi'\right) \nb\\
& &~~~ -2 k\Big(\frac{2\beta_{1}}{a^{2}} +
\frac{6\beta_{2}k}{a^{4}}
+ \frac{3g_{7}}{\zeta^{4}a^{4}}\vec\nabla^2\Big)\left(\vec\nabla^2+3k\right)\psi\nb\\
& &   ~~~~~~~~~~~~~~~ -{4\pi G a^{2}}\delta{\mu}\Big]=0,
 \eqn
  which is the same as that given in the SVW setup \cite{WM}, where
 $\delta\mu \equiv -\delta{J^{t}}/2$.
 Eq.(\ref{4.4}) represents a generalization of the 
Poisson equation of GR \cite{MW09}. 

To the first-order the supermomentum constraint (\ref{eq2}) takes the form,  
 \bq \lb{4.5}
2{\psi}' - 2kB 
= 8\pi G a {q} \,,
 \eq
which is the generalization of  the GR $0i$ constraint
\cite{MW09}, where $\delta{J}^{i} \equiv a^{-2} q^{|i}$.

On the other hand, the linearized equations (\ref{eq4a}) and (\ref{eq4b}) for the Newtonian pre-potential  
 and the gauge field  reduce, respectively, to,
 \bqn
 \lb{4.6a}
& &  \Big(\Lambda_{g} - \frac{k}{a^{2}}\Big)\Big[\vec{\nabla}^{2}B + 3 \big(\psi' + 2{\cal{H}}\psi\big) \Big]\nb\\
 & &~~~~ + \frac{2{\cal{H}}}{a^{2}}\Big[\vec{\nabla}^{2}\psi + 3 \big(2k  - a^{2}\Lambda_{g}\big)\psi\Big] = 8\pi G a \delta J_{\varphi},~~~~~~\\
 \lb{4.6b}
& & \vec{\nabla}^{2}\psi + 3k\psi = 2\pi G a^{2} \delta J_{A}.
 \eqn

The linearly perturbed dynamical equations require the calculations of the perturbed
$\left(F_{s}\right)_{ij}$ of Eq.~(\ref{eq3b}), which were given by Eq.(A1) in \cite{WM}. To avoid repeating, we shall not
write them down here, and refer readers directly to that paper. 
Then, we find that  the trace part is given by
 \bqn
\lb{4.7a}
 & & \psi'' + 2{\cal{H}}\psi'  - {\cal{F}}\psi - \frac{1}{6}
 \gamma^{ij}\delta{F}_{ij} +  {1\over 3}\left(\vec\nabla^2B'+2 {\cal H}
\vec\nabla^2B \right)\nb\\
& & ~~~~ + \frac{\bar{A}}{3a}\Big(\vec{\nabla}^{2} - 3 \Lambda_{g}a^{2} + 6 k \Big)\psi\nb\\
&& ~~~ ~
- \frac{1}{6a}\Big(2\vec{\nabla}^{2} + 3\Lambda_{g}a^{2} - 3k\Big)\delta{A}  
= 4\pi G a^{2}\delta{\cal{P}},
\eqn 
where $\delta F_{ij}=\sum g_s \zeta^{n_s}\delta (F_s)_{ij}$, with
$\delta(F_s)_{ij}$ given by Eq.~(A1) in \cite{WM},  and 
 \bqn
 \lb{4.8}
 {\cal{F}} &=& {a^2}\left(-\Lambda- {g_{1}k \over a^2}+{2\beta_1k^2
 \over a^4} +{12\beta_2 k^3 \over a^6 } \right),\nb\\
 \delta{\tau}^{ij} &=& \frac{1}{a^{2}}\left[\left(\delta{\cal P} +
2\bar{p}\psi\right)\,\gamma^{ij} + {\Pi}^{|\langle ij\rangle}\right], 
 \eqn
where the angled brackets on indices define the trace-free part:
 \bq
f_{|\langle ij \rangle} \equiv f_{|ij}-{1\over 3}
\gamma_{ij}f_{|k}{}^{|k}.
 \eq
 The trace-free part of the dynamical equations is
 \bqn
 \label{4.7b}
& & B'_{|\langle ij \rangle}+2{\cal H}B_{|\langle ij \rangle} +\delta
F_{\langle ij \rangle} - \frac{1}{a}\left(\delta{A} -\bar{A}\psi\right)_{|\langle ij \rangle}\nb\\
& & ~~~~~~~~~~~~~~~~~~~~~~  =-8\pi G a^{2} \Pi_{|\langle ij \rangle}.
 \eqn
Eqs.(\ref{4.7a}) and (\ref{4.7b}) generalize the GR $ij$
perturbed field equations \cite{MW09}.

The perturbed parts of the conservation laws (\ref{eq5a}) and
(\ref{eq5b}) give
 \bqn \lb{4.9a}
 & & \int \sqrt{\gamma} d^{3}x \Bigg\{\delta\mu' + 3{\cal H}
\left(\delta{\cal P} + \delta\mu\right) -3 \left(\bar\rho + \bar p\right){\psi}' \nb\\
& & ~  
+ \frac{1}{2a^{4}}\Bigg[\Big(a^{3} \bar{J}_{A}\Big)'  \delta{A}
+ \bar{A}\Big(a^{3}\big(\delta{A} - 3 \bar{A}\psi\big)\Big)' \Bigg]\Bigg\} =
0,\nb\\
\\
\lb{4.9b}
 & & q'+3 {\cal H}q  - a\delta{\cal P} -
{2a\over3}\left( \vec\nabla ^{2}+3k \right)\Pi 
+ \frac{1}{2}\bar{J}_{A} \delta{A} = 0, \nb\\~~
 \eqn
where $\bar{J}_{A}$ and $\bar{J}_{\varphi}$ are given by Eqs.(\ref{3.8a}) and (\ref{3.8b}).
The energy conservation equation is an integrated generalization
of the GR energy equation, and the momentum
equation generalizes the GR momentum equation
\cite{MW09}.

\section{Stability of the Minkowski Spacetime}

\renewcommand{\theequation}{5.\arabic{equation}} \setcounter{equation}{0}

It can be shown that the Minkowski spacetime, 
\bq
\lb{5.1a}
a =1, \; \;\;
 \bar{A} = \bar{\varphi}   =  k = 0,  
 \eq
 is a solution of the HMT theory, provided that
 \bq
\lb{5.1b}
   \Lambda_{g}=\Lambda    = \bar{J}_{A} = \bar{J}_{\varphi} = \bar{\rho} = \bar{p}
 = 0.
 \eq
 Then,   the linearized Hamiltonian constraint (\ref{4.4}) and the field equation (\ref{4.6a}) for $\delta{\varphi}$
are satisfied identically, while Eqs.(\ref{4.5}) and (\ref{4.6b}) yield 
\bq
\lb{5.1}
\dot{\psi} = 0 = \partial^{2}\psi, 
\eq
where $\partial^{2} \equiv \delta^{ij}\partial_{i}\partial_{j}$. These are the same as the ones obtained in GR, and lead to $\psi = 0$ 
with proper 
boundary conditions. 
It is interesting to note that in GR the equation $ \partial^{2}\psi = 0$ is obtained from the local Hamiltonian constraint,
while in the present setup it is obtained from the variation of the gauge field $A$. From this analysis, one can see clearly
the reason why $A$ is needed in order to eliminate the sprin-0 graviton. On the other hand, the trace and traceless parts of the 
dynamical equations (\ref{4.7a}) and (\ref{4.7b}) yield 
\bq
\lb{5.2}
\dot{B} = \delta{A}. 
\eq
Using the $U(1)$ gauge freedom (\ref{2.3}), without loss of generality, we can set 
\bq
\lb{5.2a}
\delta{A} = 0. 
\eq
Note that this gauge choice is consistent with our quasi-longitudinal gauge $\phi = E = 0$ \cite{WM}, because under this U(1) gauge transformation,
$E$ and $\phi$ remain the same, as one can see from Eq.(\ref{2.3}). Then, Eq.(\ref{5.2})  yields
$B = B(x)$, and  the gauge-invariant quantities $\Psi$ and $\Phi$ defined by Eq.(\ref{4.1a}) are zero, 
\bq
\lb{5.2b}
\Psi = \Phi = 0.
\eq
Therefore, the scalar perturbations of the metric vanish identically in the Minkowski background. Hence,  the spin-0 graviton is completely eliminated
in the HMT setup \cite{HMT}. 

It should be noted that $\delta\varphi$ is undetermined in the present case. However,  since $ \bar{\varphi} = 0$,  it is quite reasonable 
to assume that  its linear perturbation also vanishes 
in the Minkowski background.

\section{Conclusions}

\renewcommand{\theequation}{6.\arabic{equation}} \setcounter{equation}{0}

Recently, Horava and Melby-Thompson \cite{HMT} proposed a new  version of the HL theory of gravity, 
in which the spin-0 graviton, appearing in all the previous versions of the HL theory,  is eliminated by introducing 
a Newtonian pre-potential $\varphi$ and a local $U(1)$ gauge field $A$. Due to such an elimination, the dynamical 
coupling constant $\lambda$, which characterizes the deviation of the kinetic part of the action from that of  the 
Einstein-Hilbert, is forced to take its relativistic value $\lambda = 1$. As a result, the theory in the IR regime exhibit
many features that are quite similar to those given in GR.  

Motivated by these remarkable features, in this paper we have studied the theory in some detail by assuming the
presence of matter fields. The potential of the action has been taken to be the one constructed by SVW \cite{SVW}, which
represents the most general potential, which respects the parity and its highest order of the spatial derivatives is
six. We have first derived the Hamiltonian and super-momentum constraints, given, respectively,
by Eqs.(\ref{eq1}) and (\ref{eq2}), and then the field equations (\ref{eq4a}) and (\ref{eq4b}), respectively, 
for the Newtonian pre-potential $\varphi$ and  the local $U(1)$ gauge field $A$. The dynamical equations are 
given by Eq.(\ref{eq3}), while the conservation laws of energy and momentum are given, respectively, by 
Eqs.(\ref{eq5a}) and (\ref{eq5b}). 
 
Applying the above general formulas to cosmology, we have obtained the general modified Friedmann equation (\ref{3.6a})
and the equation (\ref{3.6b}) for the acceleration $\ddot{a}$. It is remarkable that these equations give precisely the
conservation law of energy, which takes the same form as that given in GR and can be also obtained from the conservation
law (\ref{eq5a}), despite the fact that $J_{\varphi} $ and $J_{A}$ are non-vanishing, and given, respectively, by Eqs.(\ref{3.8a})
and (\ref{3.8b}). When the spatial curvature is different from zero, terms acting as dark radiation and stiff-fluid are
present, and bouncing universe can be easily constructed from these terms. 

We have also studied the scalar perturbations of the FRW universe with any given spatial curvature, and the linearized
Hamiltonian, momentum constraints, the equations for the Newtonian pre-potential $\varphi$ and the gauge field $A$, 
the trace and traceless parts of the dynamical equations are given, respectively, by Eqs.(\ref{4.4}), (\ref{4.5}), (\ref{4.6a}), 
(\ref{4.6b}), (\ref{4.7a}), and (\ref{4.7b}), while the conservation laws of energy and momentum are given, respectively,
by Eqs.(\ref{4.9a}) and (\ref{4.9b}).

Applying these formulas to the Minkowski background, we have shown explicitly that the metric scalar perturbations vanish
identically, that is, the spin-0 graviton appearing in all the previous versions of the HL theory is eliminated in the current
HMT setup.   
 
Since the Newtonian pre-potential $\varphi$ and the gauge field $A$ have no contributions to the vector and tensor
perturbations, the corresponding linear perturbations are given precisely by the same equations as those recently presented
in \cite{Wang} in the SVW setup. In particular,  in the Minkowski background   vector perturbations also vanish, although  the tensor
perturbations in general do not \cite{Wang}. These two non-vanishing components represent the massless spin-2 gravitons,
 which are exactly the same as   those found  in GR.


~\\{\bf Acknowledgements:}  The work of AW was supported in part by DOE  Grant, DE-FG02-10ER41692.


\end{document}